\documentclass[twocolumn,11pt]{article}
\usepackage{graphicx}         % Per l'inserimento di immagini
\usepackage[T1]{fontenc}      % Migliore gestione dei caratteri
\usepackage[utf8]{inputenc}   % Per caratteri speciali e accenti
\usepackage{natbib}
\usepackage{hyperref}

\usepackage[
    margin=1.5cm,        % Imposta tutti i margini a 1.5 cm
]{geometry}

\title{SHARP\thanks{SHARP website http://sharp.brera.inaf.it}: Beyond JWST \\
Revealing the galaxy birth and growth with the 
resolution of the ELT}
\author{
\parbox{\textwidth}{ % Usa l'intera larghezza del testo
\centering % Centra l'intero blocco
P. Saracco$^1$, P. Conconi$^1$, C. Arcidiacono$^2$, H. Mahmoodzadeh$^1$, I. Di Antonio$^3$,
E. Portaluri$^3$,
P. Franzetti$^4$, A. Gargiulo$^4$, E. Molinari$^1$, J. M. Alcal\'a$^5$, S. Bisogni$^4$,
R. Bonito$^6$, E. Bortolas$^2$, M. Cantiello$^3$, E. Cascone$^5$, V. Cianniello$^5$, E. M. Corsini$^7$,
F. D’Ammando$^8$, E. Dalla Bont\`a$^7$, M. Dall'Ora$^5$, V. De Caprio$^5$, G. De Lucia$^9$,
B. Di Francesco$^3$, G. Di Rico$^3$, C. Eredia$^5$, M. G. Guarcello$^6$, L. Izzo$^5$, F. La Barbera$^5$,
M. Lippi$^{10}$, M. Longhetti$^1$, A. Longobardo$^{11}$, C. Mancini$^4$, M. Mirabile$^3$, E. Piconcelli$^{12}$,
A. Pizzella$^7$, L. Podio$^{10}$, L. Prisinzano$^6$, C. Tortora$^5$, G. Vietri$^4$ and H.-F. Wang$^2$
}
}

\date{}
\begin{document}
\maketitle
\begingroup
\renewcommand\thefootnote{\textsuperscript{\arabic{footnote}}} % Usa numeri arabi per le note
\footnotetext[1]{INAF - Osservatorio Astronomico di Brera, Milano, Italy}
\footnotetext[2]{INAF - Osservatorio Astronomico di Padova, Italy}
\footnotetext[3]{INAF - Osservatorio Astronomico d'Abruzzo, Teramo, Italy }
\footnotetext[4]{INAF - IASF, Milano, Italy}
\footnotetext[5]{INAF - Osservatorio Astronomico  di Capodimonte, Napoli, Italy}
\footnotetext[6]{INAF - Osservatorio Astronomico di Palermo, Palermo, Italy} 
\footnotetext[7]{Università degli Studi di Milano-Bicocca, Milano, Italy}
\footnotetext[8]{INAF - IRA, Bologna, Italy}
\footnotetext[9]{INAF - Osservatorio Astronomico di Trieste, Trieste, Italy }
\footnotetext[10]{INAF - Osservatorio Astrofisico di Arcetri, Firenze, Italy }
\footnotetext[11]{INAF - IAPS, Roma, Italy}
\footnotetext[12]{INAF - Osservatorio Astronomico di Roma, Roma, Italy}
% ... Aggiungi tutte le altre affiliazioni che ti servono
\endgroup

\begin{abstract}
A deep understanding of the life-cycle of galaxies, particularly those of high mass,
requires clarifying the mechanisms that regulate star formation (SF) and its abrupt shutdown (quenching), often capable of stopping SF rates of hundreds of solar masses per year.
What initially triggers quenching, and what sustains the quiescent state thereafter, especially given the frequent presence of large gas reservoirs or even massive gas inflows,
are unsolved key issues.
Ultimately, the crucial connection between the galaxy life-cycle and the surrounding Intergalactic (IGM) and Circumgalactic (CGM) Medium remains largely unclear.
Addressing these issues requires studying star formation, chemical enrichment, and quenching
homogeneously up to high redshift. 
The upcoming AO-assisted Extremely Large Telescope (ELT), will deliver sharper and deeper data than the JWST. 
SHARP is a concept study for a near-IR (0.95–2.45 $\mu$m) spectrograph designed to fully exploit the capabilities of ELT.
Designed for multi-object slit spectroscopy and multi-Integral Field spectroscopy, 
SHARP points to achieve angular resolutions ($\sim$30 mas) far superior to NIRSpec@JWST 
(100 mas) to decipher and reconstruct the life-cycle oa galaxies. 

\end{abstract}

\section{Massive galaxies} 
According to the current hierarchical model of galaxy formation, Dark Matter (DM) 
halos form the seeds of the first galaxies, which grow over time via baryon accretion, 
star formation, and merging \citep[e.g.,][]{springel05}. 
In this paradigm, massive galaxies assemble late, becoming common in the local Universe 
but increasingly rare and eventually absent in the earliest cosmic epochs 
\citep[e.g.,][]{delucia13}.
%Therefore, the search for massive galaxies at high redshifts has long been a powerful 
%test of the hierarchical model \citep[e.g.,][]{thompson99,saracco03}. 
Deep observations over the past decade have uncovered numerous massive 
galaxies at high redshift, often belonging to overdensities. 
In practice, they appear to be too massive, too numerous, and formed too early 
on cosmic timescales compared to theoretical expectations, significantly challenging our 
current understanding of galaxy formation physics \cite[e.g.,][]{boylan23,chworowsky24,carnall24,glazebrook24}.

\paragraph{Star formation} -
JWST has detected massive quiescent galaxies up to $z$$\sim$4-5, 
when the Universe was younger than 1.2 Gyr \cite[e.g.,][]{carnall24}. 
These include both galaxies whose stellar population is maximally old, that is, formed 
few million years after the Big Bang (i.e., $z$$\sim$11-12) and galaxies whose stars are 
much younger, having formed few million years before observation (i.e., $z$$<$5) \cite[e.g.,][]{tanaka19,saracco20apj,deugenio20,antwi-danso25,carnall24,glazebrook24}. 
Given the high stellar masses (log(M*/M$\odot$)$\sim$11) and the few million years 
available to form them, the resulting star formation rates (SFR) must be higher 
than hundreds solar masses per year.
These values are particularly challenging under the conditions of the early Universe: 
the efficiency of star formation is expected to be reduced by several factors, 
including the lower cooling efficiency of pristine hydrogen, the higher UV radiation 
density and pressure, and the fact that Dark Matter (DM) halos of sufficient hosting 
mass were possibily not yet formed \cite[e.g.,][]{boylan23,glazebrook24}.

\paragraph{Metallicity and enrichment} - 
Massive galaxies, whether they host old or young stellar populations, are characterized 
by solar metallicity or higher, e.g., [Fe/H]$\sim$0.02 \citep[][]{glazebrook24} and
[Z/H]$>$0.15 \cite[][]{saracco20apj}.
Such high metallicity values further challenge theories of chemical enrichment, since
such levels require times ($>$1 Gyr) apparently inconsistent with those of the 
star formation times deduced for these galaxies.

\paragraph{Quenching and quiescence} - 
Since both massive high-redshift galaxies with young stellar populations and those with 
old stellar populations appear quiescent, with virtually no ongoing star formation, 
an extremely efficient shutdown mechanism must have taken place, capable of abruptly 
halting star formation rates of hundreds of solar masses per year.
Quenching mechanisms alternative to AGN outflows are needed, since signs of outflow 
(e.g., asymmetric absorption lines) are seen in very few high-redshift post-starburst 
galaxies, and outflows seem to be not efficient in removing gas, the way to abruptly 
halt star formation \citep{concas22}. 

Also, most of the massive galaxies must remain quiescent, without experiencing further 
significant episodes of star formation, to prevent them from exceeding the mass of (and 
appearing younger then) local counterparts. 
This assumption holds despite the presence in many of them of a significant quantity 
of residual gas and/or massive gas inflows \cite[e.g.,][]{belli24,bevacqua25}. 
The reason why these galaxies fail to ignite subsequent star formation in such 
gas-rich environments remains an open question.

\paragraph{Mass growth and hierarchical assembly} -
Indeed, massive, high-redshift ($z$$\sim$3-4) galaxies that host old 
($\sim$1.5-2 Gyr) stellar populations may have accumulated their mass through 
the merger of smaller, primordial galaxies, since time could be sufficient.
Their presence could be still consistent with the hierarchical paradigm, even if
this scenario still faces the challenge of explaining the high star 
formation rate required even if spread among multiple galaxies.
On the contrary, the presence of massive galaxies at high redshift hosting young 
($\sim$0.5 Gyr) stellar populations, i.e., formed few million years before the
observations, 
challenges hierarchical paradigm because the stellar mass must have formed in situ
\citep[e.g.,][]{puskas25}, 
given that there is insufficient time for assembly through the merging of individual 
subunits \citep{boylan23}.

\section{Observational needs: SHARP}
High-redshift massive galaxies pose significant challenges to current knowledge 
due to evidence of inexplicably high star formation rates, fast chemical enrichment, 
and extremely rapid quenching mechanisms.
Addressing these issues requires studying star formation, chemical enrichment, and quenching
homogeneously up to high redshift. 
This necessitates resolving physical scales comparable to Giant Molecular Clouds (GMCs).
GMCs are the fundamental star-forming units, primary sites of metal production, and presumed 
cradles of globular clusters and possibly the first Pop III systems. 
With typical sizes of 150-250 pc, containing over $10^6$ M$\odot$ of molecular gas, 
an angular resolution of $\sim$ 30-35 mas is required to sample these scales across 
cosmic time.
Therefore, Near-IR spectroscopic observations on multiple sources, assisted by MCAO 
and coupled with a pixel scale of $\sim$ 30 mas/pixel, are mandatory. 
\begin{figure*}
\begin{center}
\includegraphics[width=12.0cm]{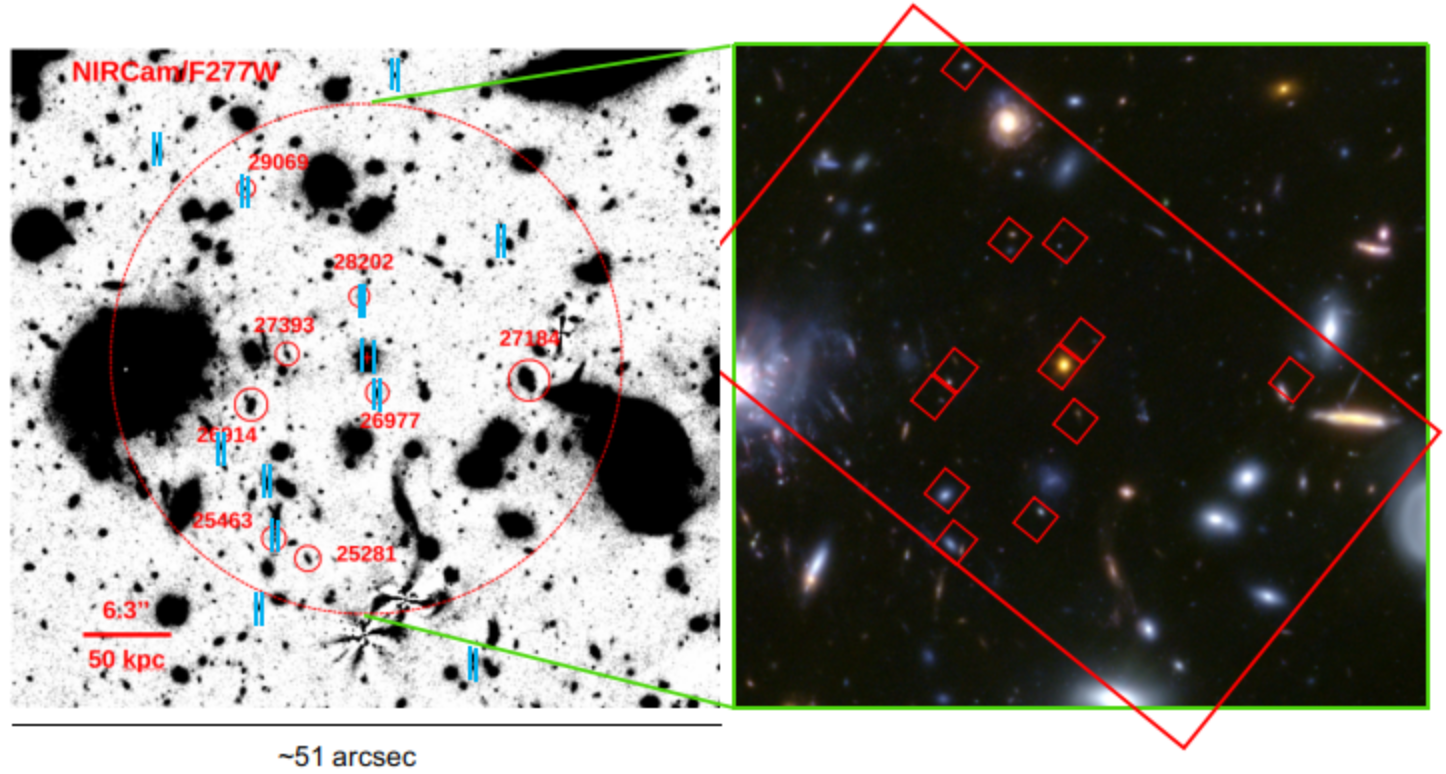}
\end{center}
\caption 
{\label{fig:jades} Left - NIRCam image of the field  (51"x51") centered on galaxy GLASS-180009, adapted from \cite{bevacqua25} (Fig. 5).
The red small circles mark galaxies with similar redshift within a radius of 
about 150 kpc (large red circle).
The FoV of NEXUS ($72'' \times 72''$) fully encompasses the GLASS-180009 field. 
The light-blue small double-lines represent the slits of NEXUS, whose subtended field 
can be rotated thanks to the inversion prisms \citep{saracco24}
Right - Zoom-in ($38'' \times 38''$) composite JWST image centered on GLASS-180009. 
The big red rectangle marks the area ($\sim 20.5'' \times 40''$) probed by the 
12 probes (small red squares, $\sim 1.7'' \times 1.5''$ each) of VESPER.
}
\end{figure*}

These measurements are currently beyond the capabilities of JWST due to its 
limited angular resolution (0.1", or $\sim$800 pc at $z$$\sim$1-4).  
Furthermore, they challenge the existing ELT spectrographs. 
Indeed, the spectroscopic mode of MICADO, single-slit and, possibily, single IFU as well
as HARMONI (single IFU, $\sim$3"$\times$4") are constrained by a their limited field 
of view and the inability to observe multiple sources simultaneously. 
{\it The crucial issue is that the full potential of a large, uniformly corrected field at the 
ELT diffraction limit, as provided by an MCAO unit like MORFEO, is not exploited by the ELT spectrographs currently planned to be supported by MORFEO.}

Addressing this gap is the objective of SHARP.

SHARP  \citep{saracco24,mahmoodzadeh25} 
is composed of two main units: NEXUS, a Multi-Object Spectrograph and VESPER, 
a multi-object Integral Field Unit. NEXUS is fed by a configurable 
slit system capable of deploying $\sim$30 slits  over an AO corrected field of $\sim$1.2’x1.2’. The pixel size is 35 mas/pixel. VESPER is a modular system 
composed of two modules of 6 probes each totaling 12 probes deployable 
over an AO corrected area of 20”x40”. Each probe has a FoV~1.7”x1.5” sliced 
at 31 mas.

\section{GLASS-180009 ($z$$\sim$2.66)}
In Fig. \ref{fig:jades} (left panel) it is shown the NIRCam image centered on 
galaxy GLASS-180009, as adapted from Fig. 5 in \cite{bevacqua25}. 
GLASS-180009 at $z$$\sim$2.66 serves as a typical case study encompassing 
multiple astrophysical features, demonstrating the  capacity of SHARP to meet complex 
observational requirements.

GLASS-180009 is old ($\sim$ 1.7 Gyr), massive ($\sim$ 4 $\times$$10^{10}$ M$\odot$), 
and quiescent (SFR$<$0.2  M$\odot$/yr) \citep{marchesini23}. 
Its stellar mass formed rapidly at $z$$\sim$11 \citep{bevacqua25}.
The galaxy exhibits a detected neutral gas inflow (M$_{gas}$$\sim$$10^8$ M$_\odot$; 
rate $\sim$19 M$_\odot$/yr) via the redshifted NaI doublet \citep{bevacqua25}. 
Despite this gas reservoir, the galaxy remains quiescent, and the nature of the 
inflow from the IGM, cosmic filaments, or nearby companions in the likely overdensity region is unknown.
SHARP's capabilities allow for the simultaneous study of the galaxy and its environment, 
which is not possible with NIRSpec at JWST (FoV$\sim$3"$\times$3" with 0.1" sampling).

An MCAO unit like MORFEO at the ELT uniformly corrects for atmospheric turbulence 
a field even larger than that surrounding GLASS-180009 shown in the left panel 
of Fig.  \ref{fig:jades}.
This, in principle, allows for probing the nature of the inflowing gas through
observations of GLASS-180009 and the surrounding regions, searching for signs IGM 
or gas streams, as well as studying the properties of the surrounding galaxies.
\begin{figure}
\begin{center}
\includegraphics[width=8.0cm]{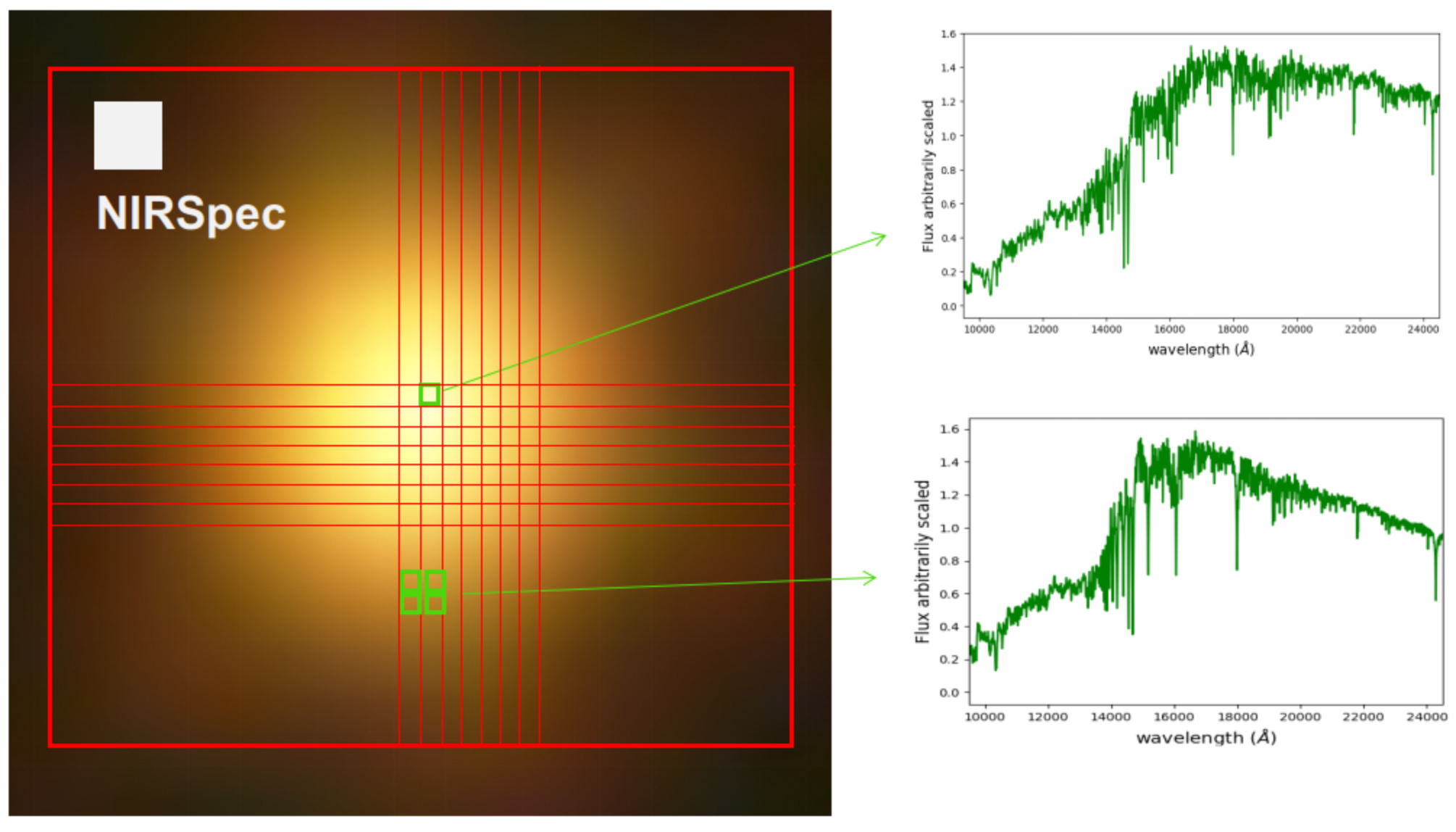}
\end{center}
\caption 
{ \label{fig:jades2} Left - Zoom in of the composite JWST image centered on GLASS-180009. 
The big red square is the area ($\sim1.7'' \times1.5''$) of a single FS of VESPER.
The thin red lines schematically represent the slicing at 0.031".
Highlighted in green are the central spaxel with a corresponding simulated spectrum representing a SSP 1.75 Gyr old and the sum of four spaxel in the outer region
with a corresponding SSP of 0.9 Gyr old.
The gray square represents the pixel size of NIRSpec (0.1").
}
\end{figure}

MOS observations would be necessary to identify all galaxies at  
comparable redshift, defining their integrated properties, derive their
kinematics, thereby defining the extension of the overdensity region 
(if any), characterizing the properties of its members, and ultimately constraining 
the properties of the overdensity itself.
NEXUS, the MOS unit of SHARP, would allow us to carry out such observations targeting some 
surrounding galaxies (blue double-lines).
Target galaxies can be aligned to the slits thank to the inversion prisms \citep[see][]{saracco24}.
This represents a unique step forward in exploiting the high angular resolution 
over a wide field typical of MCAO systems, without which it would not be possible 
to obtain spatially resolved information for so many galaxies simultaneously.

In the right-hand panel of Fig. \ref{fig:jades} it is shown a square region of 
about $38'' \times 38''$ centered on GLASS-180009. 
Superimposed is the area ($\sim 20.5'' \times 40''$) probed by the 12 probes of VESPER,
the multi-IFU unit of SHARP. 
The 12 probes (1.7"x1.5" each) are arranged to sample both the region close to the galaxy and
some of the surrounding galaxies.
This observation simultaneously probes the nature of the inflow, its relationship 
with the IGM (if present), and/or with surrounding galaxies.

Fig. \ref{fig:jades2} shows the field of one probe of VESPER ($\sim$1.7"$\times$1.5")
centered on GLASS-180009.
The effective diameter of the galaxy (enclosing 50\% of the light) is about 
2 kpc ($\sim$0.25"), sampled by VESPER at about 250 pc (0.031").
The spatially resolved information, on these scales for the galaxy 
GLASS-180009, allows us to investigate the conditions  of quiescence 
despite the available gas. 
Crucially, it enables the determination of gradients in the stellar population 
properties (age, metallicity, and possibly the IMF) and the kinematics, 
ultimately permitting the reconstruction of the galaxy's mass assembly history
via kinematic mapping.

\section{SHARP complementarity}
SHARP is highly complementary to other ELT spectrographs. 
For instance, the fiber-fed MOSAIC \citep{pello24} is limited to 1.8 $\mu$m 
(precluding detection of many atomic features at $z>2.6$) and features a
lower angular resolution ($\sim 0.2''$ fiber diameter). 
Nevertheless, MOSAIC is complementary by enabling large-scale surveys across a much wider field of view than the MORFEO-corrected area.

Similarly, ANDES\citep{marconi24} (also limited to $\lambda$$<$1.8 $\mu$m) 
will be complementary as a single-object spectrograph, offering an extremely 
high spectral resolution ($R=100000$), essential for detailed analysis of chemical abundances and kinematics on the IGM and exoplanet atmospheres.
% --- Inizio Blocco per Bibliografia Compatta ---

\setlength{\parskip}{0pt}       % Assicurati che non ci sia spazio tra i paragrafi
\setlength{\itemsep}{0pt}       % Riduce lo spazio tra le singole voci (come hai provato)
\setlength{\parsep}{0pt}        % Riduce lo spazio tra paragrafi all'interno delle voci

% Questo comando agisce sullo spazio verticale prima e dopo la bibliografia
% e su quello tra le voci, ed è spesso il più efficace.
\let\oldthebibliography\thebibliography
\let\endoldthebibliography\endthebibliography
\renewenvironment{thebibliography}[1]{
  \begin{oldthebibliography}{#1}
    \setlength{\itemsep}{0pt}     % Di nuovo, per sicurezza
    \setlength{\parskip}{0pt}     % Di nuovo, per sicurezza
    \setlength{\baselineskip}{12pt} % Opzionale: controlla la linea base
    \small                        % Opzionale: usa font più piccolo
}{\end{oldthebibliography}}

% --- Fine Blocco per Bibliografia Compatta ---
\small
\section*{Acknowledgments}
The SHARP team acknowledges support by Bando Ricerca Fondamentale INAF 2022, 
Techno-Grant "SHARP" - 1.05.12.02.01 and Bando Ricerca Fondamentale INAF 2024, 
Large-Grant "SHARP" - 1.05.24.01.01.
\vskip-1truecm
\bibliographystyle{aa} 
% Loading bibliography database

\bibliography{sharp.bib}
%\end{multicols}}

\end{document}